\documentclass[twocolumn]{revtex4}
\def\address{\affiliation}
\usepackage[dvips]{graphicx}
\begin{document}


\title
[
Anomalous Pd substitution effects in NaCo$_{2-x}$Pd$_x$O$_4$
]
{
Anomalous Pd substitution effects in 
the thermoelectric oxide NaCo$_{2-x}$Pd$_x$O$_4$
}

\author{
R Kitawaki and I Terasaki\footnote[1]
{Corresponding author. E-mail address: terra@waseda.jp}
}

\address{
Department of Applied Physics, Waseda University,
Tokyo 169-8555, JAPAN
}

\date{\today}

\begin{abstract}
We prepared a set of polycrystalline samples of the thermoelectric oxide  
NaCo$_{2-x}$Pd$_x$O$_4$ ($x$= 0, 0.05, 0.1 and 0.2), and 
investigated the Pd substitution effects on transport phenomena.
The effects are so drastic that only 5-10\% Pd ions reduce
the resistivity and the Seebeck coefficient to one-third 
of the values for $x=0$, 
and increase the magnitude of the Hall coefficient by three times.
A semi-quantitative analysis has revealed that the $x=$0.2 sample
has much smaller effective mass and 
carrier concentration than the $x=$0.05 sample.
This is difficult to explain within a rigid-band picture,
and is qualitatively consistent with a strong-correlation picture
applied to the Ce-based heavy fermion systems.
\end{abstract}

\maketitle
\section{Introduction}
A thermoelectric material converts heat into electricity and vice versa 
through the Seebeck and Peltier effects,
which is characterized by a low resistivity ($\rho$), a large
Seebeck coefficient ($S$), and a low thermal conductivity ($\kappa$).
Thermoelectric devices are attracting renewed interests in recent years, 
because they are direct energy conversion devices without any wastes, 
and can work for a long time without maintenance. 

Terasaki {\it et al}. \cite{terasaki}  discovered 
high thermoelectric properties in NaCo$_2$O$_4$ in 1997.
A single crystal of this material showed low resistivity
(200~$\mu\Omega$cm) and large Seebeck coefficient 
(100~$\mu$V/K) at room temperature.
Fujita \textit{et al}. \cite{fujita} reported a low value of $\kappa$ 
(50 mW/cmK) for single-crystal NaCo$_2$O$_4$ at 800 K, 
which strongly suggests that NaCo$_2$O$_4$ 
is a potential thermoelectric material at high temperature.

In spite of low thermoelectric performance for conventional oxides, 
the thermoelectric properties of NaCo$_2$O$_4$
are exceptionally high.
Thus a central issue is to elucidate the mechanism of 
the thermoelectric properties of NaCo$_2$O$_4$.
We have proposed that the strong (electron-electron) correlation plays 
an important role in this compound, as is similar
to the case of the Ce-based intermetallic compounds
(so-called valence fluctuation/heavy-fermion systems) \cite{terasaki2}. 
Koshibae {\it et al}. \cite{koshibae}
evaluated $S$ induced from the correlated electrons 
to be 150 $\mu$V/K in the high-temperature limit 
by using the extended Heikes formula,
which roughly agrees with experiments.
On the other hand, Singh \cite{singh} predicted a large value of 
$S$=100 $\mu$V/K at 300 K on the basis of the band calculation.

We have studied various substitution effects to examine 
whether or not the band picture is broken down by the correlation.
Upon the Na-site substitution, the transport properties were 
rather insensitive, where the Ca substitution reduced 
carrier concentration slightly \cite{kawata}.
Most of the impurities substituted for Co acted as 
a strong scatterer, and increased resistivity with 
a strong upturn at low temperatures \cite{ict2000}.
Exceptions were found in Cu- and Pd-substitutions for Co.
The Cu substitution improved the thermoelectric 
performance \cite{Cu}, which was successfully explained in 
analogy to the Ce-based intermetallic compounds \cite{Cu2}.  
The Pd substitution is another anomalous case,
where it decreases both the resistivity and 
the Seebeck coefficient.
In this paper we report on a semi-quantitative analysis
of the Pd substitution effect in NaCo$_{2-x}$Pd$_x$O$_4$,
which strongly supports our strong correlation picture
rather than the simple band picture.

\section{Experimental}
Polycrystalline samples of Na$_{1.2}$Co$_{2-x}$Pd$_x$O$_4$ 
($x$=0, 0.05, 0.1 and 0.2) were prepared by a solid-state reaction. 
Stoichiometric amounts of Na$_2$CO$_3$, Co$_3$O$_4$ and PdO 
were mixed, and the mixture was calcined at 860$^{\circ}$C for 12~h in air. 
The product was finely ground, pressed into a pellet, and sintered at 
920$^{\circ}$C for 12~h in air.
Since Na tends to evaporate during calcination, we added 20\% excess Na.
We expected samples of the nominal composition of 
Na$_{1.2}$Co$_{2-x}$Pd$_x$O$_4$
to be NaCo$_{2-x}$Pd$_x$O$_4$.

The resistivity was measured by a four-terminal method from 
4.2 to 300 K in a liquid He cryostat. 
The Seebeck coefficient was measured using a steady-state technique 
with a typical temperature gradient of 0.5~K/cm 
from 4.2 to 300 K in a liquid He cryostat.  
The thermopower of the voltage leads was carefully subtracted.
The Hall coefficient ($R_H$) was measured in a closed refrigerator
from 15 to 200 K. 
A CERNOX resistive thermometer was placed at 45 cm above 
the magnet core, which successfully suppressed the 
magnetoresistance of the thermometer to keep the accuracy of
the measured temperature within 0.01\% at 7 T.
An AC-bridge nano-ohmmeter was used to measure the resistivity 
by sweeping magnetic field from -7 to 7 T in 20 minutes 
at constant temperatures.
An unwanted signal occurring from a misalignment of the 
voltage pads was carefully removed by subtracting 
negative-field data from positive-field data.
The Hall voltage was linear in magnetic field,
and $R_H$ was determined by the data at $\pm$7 T.

\begin{figure}[b]
 \begin{center}
  \includegraphics[width=7.8cm,clip]{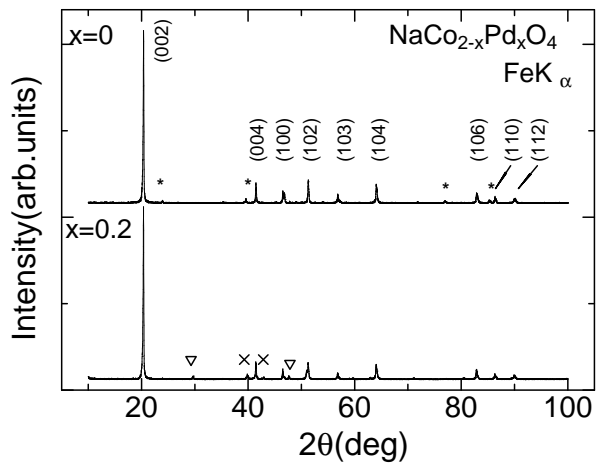}
 \end{center} 
 \caption{
 The X-ray diffraction patterns of polycrystalline 
 NaCo$_{2-x}$Pd$_x$O$_4$.
 }\label{f1}
\end{figure}

The X-ray diffraction (XRD) of the samples was measured 
using a standard diffractometer with Fe K$_\alpha$ radiation 
as an X-ray source in the $\theta -2\theta$ scan mode. 
Figure \ref{f1} shows typical XRD patterns of 
NaCo$_{2-x}$Pd$_x$O$_4$.
Almost all the peaks are indexed as the $\gamma$ phase \cite{jansen}.
For $x=0$, a small amount of Co$_3$O$_4$ 
indicated by `$\ast$' appears.
For $x$=0.2, a tiny trace of PdO indicated by `$\times$' and
an unidentified phase indicated as `$\bigtriangledown$'
appears instead of the peaks of Co$_3$O$_4$.
Although Pd did not fully substitute for Co in a strict sense, 
the volume fraction of the impurity phases (3\%)
is small enough to retain high signal-noise ratio of the XRD pattern.
In the next section we will see a systematic evolution of the transport 
parameters with the Pd substitution, which allows us to 
conclude that Pd in NaCo$_{2-x}$Pd$_x$O$_4$ surely modified the
electronic states of the host.

\section{Results and Discussion} 
\begin{figure}
 \begin{center}
  \includegraphics[width=7.5cm,clip]{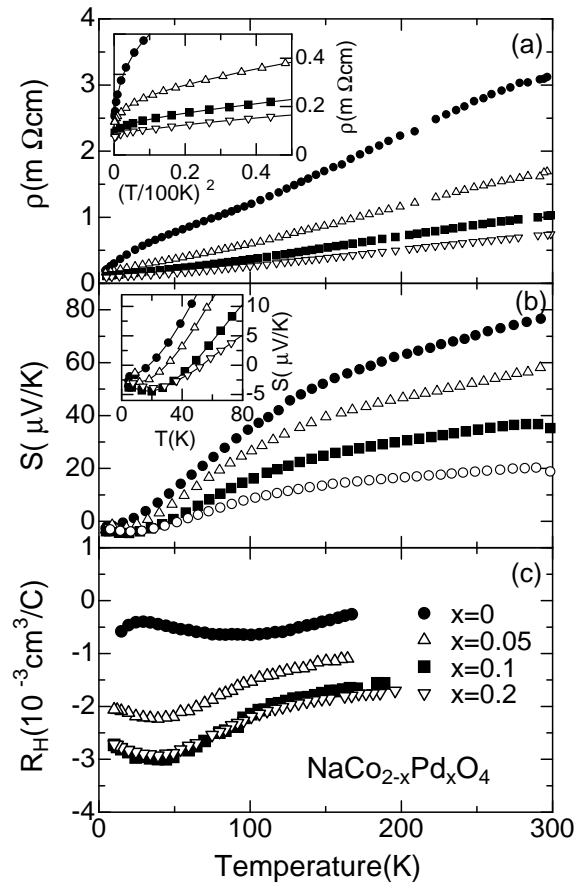}
 \end{center}
 \caption{
 (a) The resistivity ($\rho$), (b) the Seebeck coefficient ($S$) and 
 (c) the Hall coefficient ($R_H$)
 of polycrystalline NaCo$_{2-x}$Pd$_x$O$_4$ 
 plotted as a function of temperature ($T$).
 }\label{f2}
\end{figure}

Figure \ref{f2}(a) shows the resistivity of the prepared samples of 
NaCo$_{2-x}$Pd$_x$O$_4$ plotted as a function of temperature ($T$). 
The resistivity decreases systematically with the Pd content $x$,
whose magnitude decreases from 3 m$\Omega$cm for $x$=0
to 0.7 m$\Omega$cm for $x$=0.2 at room temperature.
This is a quite unusual substitution effect.
NaCo$_2$O$_4$ is a layered oxide consisting of the conductive CoO$_2$
layer and the insulating Na layer, where an impurity in the conductive layer 
would normally act as a scattering center.
Thus one can expect that the impurity induces a residual
resistivity in conventional metals, and in fact the Mn- Fe- Ru- and 
Rh-substitutions dramatically increase the residual resistivity
in NaCo$_2$O$_4$ \cite{ict2000}.

The inset of figure \ref{f2}(a) shows the low-temperature resistivity
of the same samples as a function of $(T/100~ {\rm K})^2$, 
where no resistivity upturn is seen.
This indicates that the Pd substitution makes no localization down to
4.2 K, and that the scattering cross section of Pd  is negligibly small.
We should further note that all the resistivities strongly depend on
temperature down to 4.2 K, which suggests that the carriers are
predominantly scattered through the electron-electron interaction,
not through the electron-phonon interaction.
Thus the electron-electron correlation still dominates in the
most conductive sample of $x=$0.2, 
the low-temperature resistivity for which is roughly proportional to $T^2$
that is usually expected for the electron-electron scattering.

Figure 2 (b) shows the Seebeck coefficient of 
NaCo$_{2-x}$Pd$_x$O$_4$ plotted as a function of temperature. 
As is similar to the resistivity, the Seebeck coefficient 
decreases with $x$, where $S$ for $x=0.2$ (20 $\mu$V/K at 300 K) 
is about quarter of that for $x=0$.
This is another quite anomalous substitution effect.
Since the Seebeck coefficient is independent of scattering time
in the lowest order approximation, it is insensitive to
the disorder and/or impurities in usual cases. 
However, the substituted Pd not only decreases the magnitude of $S$,
but also modifies the temperature dependence.
As is shown in the inset of figure \ref{f2}(b),
$S$ goes to negative at low temperatures.
We do not understand the mechanism of the sign change at present.
Suffice it to say that similar sign change is observed in 
Ce$M_2$Si$_2$ ($M$=Au, Pd, Rh and Ru)\cite{amato},
whose origin is not fully understood either.

Figure 2 (c) shows the temperature dependence of 
the Hall coefficient of NaCo$_{2-x}$Pd$_x$O$_4$. 
The sign is negative for all the samples,
and the temperature dependence  becomes weaker with $x$.
The magnitude of $R_H$ significantly increases 
from 5$\times$10$^{-4}$cm$^3$/C
for $x=0$ to 3$\times$10$^{-3}$cm$^3$/C for $x=0.2$ at 15 K.
We should emphasize that a value of 3$\times$10$^{-3}$cm$^3$/C
is considerably large, implying that the k-space volume surrounded 
by the Fermi surface is sufficiently small.
The Hall coefficient for a two-band model is written as 
$R_H=(n_p-n_e)/e(n_p+n_e)^2$, where $n_p$ and $n_e$ are 
the carrier concentrations for the hole band and the electron band,
respectively (for simplicity, the same mobility is assumed).
Since this expression clearly shows $|1/eR_H| > n_e$,
the electron concentration is smaller than 2-3$\times 10^{21}$ cm$^{-3}$.
We can also point out that $n_e$ and $n_p$ should be of the order of
$10^{21}-10^{22}$ cm$^{-3}$, considering that the formal valence of Co
does not change so much from +3.5.
Thus we conclude that the minority carrier concentration 
must be small for $x=0.2$, and that the electric conduction 
is dominated by a single band.
We note that  $|R_H|$ for $x=0.2$ is as large as  $|R_H|$
of optimally-doped high-temperature superconductors,
which has been analysed with a single band \cite{hall}.
$R_H$ for $x=0.2$ is as weakly dependent on temperature as 
$R_H$ for high-temperature superconductors, 
which is further consistent with the single-band picture,
because a temperature-dependent $R_H$ is often due to 
different mobilities in multi-bands.

Now we will discuss the Pd-substitution effect more quantitatively.
As mentioned above, the large value of $|R_H|$ is likely to allow us to 
apply a single-band picture,
although the signs of $S$ and $R_H$ are different.
Then we can evaluate the carrier concentration ($n$), 
the effective mass ($m$) and the scattering time ($\tau$) 
for the single band.
In the lowest order approximation, $\rho$, $R_H$ and $S$ are
expressed as functions of $n$, $m$ and $\tau$ as
\begin{eqnarray}
\rho  &=& \frac{m}{ne^{2}\tau}  \label{r}\\
|R_{H}| &=& \frac{1}{ne}  \label{rh}\\
|S|     &=& \frac{{\pi}k_B^2}{2{\hbar}^{2}d_{c}e}\frac{m}{n}T \label{s} 
\end{eqnarray}
where $d_c$(=0.54 nm) is the inter-layer spacing,
and $e(>0)$ is the unit charge. 
For the expression of $S$, we assumed the two-dimensional 
Fermi surface \cite{mandal}.

Unfortunately  equations (\ref{r})-(\ref{s}) are valid only 
in the limited temperature range,
where $T$-linear $S$ and $T$-independent $R_H$ are expected.
Thus we conclude that the data below 100 K is unsuitable for the analysis.
The upper limit is set to be 160 K,
the highest measured temperature for $R_H$ of $x=0$.
Consequently we have chosen 100 and 160 K as 
two representative temperatures.

\begin{figure}
 \begin{center}
  \includegraphics[width=6cm,clip]{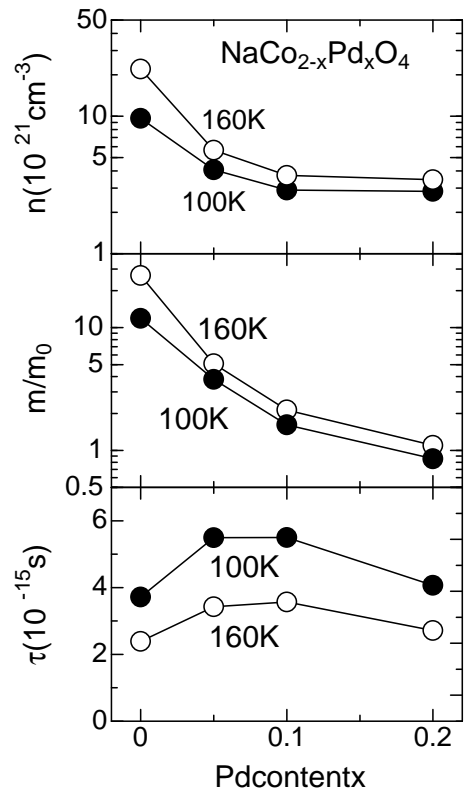}
 \end{center}
 \caption{
 The carrier concentration ($n$), the effective mass
 normalized by the free electron mass ($m/m_0$) and 
 the scattering time ($\tau$) of polycrystalline 
 NaCo$_{2-x}$Pd$_x$O$_4$ plotted as the function of Pd content $x$.
 }\label{f3}

\end{figure}

Figure 3 shows thus evaluated $n$, $m/m_0$ and $\tau$ 
using equations (\ref{r})-(\ref{s}) and the data in figure 2, 
where $m_0$ is the free electron mass.
In spite of the rough assumptions, 
the parameters are reasonably evaluated: 
$n$ and $m$ are found to be essentially 
independent of temperature, and is highly dependent on $x$.
On the other hand, $\tau$ is highly dependent on temperature,
but is weakly dependent on $x$.
These are indeed what we expect in a simple metal of a single band.
Apparent $T$-dependence in $m$ and $n$ for $x=0$
is due to the fact that the single-band analysis 
is not guaranteed for $x=0$ because of the small $|R_H|$.
Nonetheless the three parameters are systematically
varied from $x=0.2$ down to $x=0$,
which strongly suggests that the single-band picture
is more or less valid for all the samples.
In particular, we think that the data for $x\ge0.05$ are reasonably evaluated.

Let us take a closer look at the $x$ dependence of $n$, $m$ and $\tau$.
Reflecting that the substituted Pd does not cause
the residual resistivity,  $\tau$ is evaluated to be essentially 
independent of $x$, which indicates that the scattering cross section of
Pd is  negligibly small.
The magnitude is of the order of $10^{-15}$ s,
which is as large as $\tau$ of usual metals.
The evaluated magnitudes of 
$n$ and $m$ for $x=0$ are satisfactory,
although the single-band analysis is less reliable than for $x>0$.
A value of $m/m_0=10-30$ is consistent with 
the specific-heat measurement \cite{ando},
and $n=$10$^{22}$cm$^{-3}$
is of the same order of the value estimated from 
the formal valence of Co ($3.5+$).
An important finding is that 
the $x$ dependence of $m$ and $n$ is surprisingly large.
In going from $x=0.05$ to $x=0.2$, $n$ decreases by 1.5-2 times, 
and $m$ decreases by 4-5 times.
As a result, $n/m$ is found to increase by 3 times from $x$=0.05 to 0.2,
which is the origin for the decrease in  $\rho$ and $S$ with $x$.

We should note that the rapid decrease in $m$ is seriously incompatible 
to the band picture.
According to the band structure by Singh \cite{singh},
the band dispersion near the Fermi energy is smooth 
without any singularities.
Thus the band mass is unlikely to change by 4-5 times
in the rigid-band picture.
Furthermore the $a_{1g}$ band (responsible for the large S)
forms the cylindrical Fermi surface,
which makes the density of states at the Fermi energy  
nearly independent of $x$ in the rigid-band picture.
Thus the rapid decrease in $n$ with $x$ is also incompatible to 
the band picture.

On the contrary, the strong correlation scenario can, at least 
qualitatively, explain the rapid change in $n$ and $m$.
In the previous paper, we compared the physical properties 
of the Cu substituted samples with the Ce-based compounds \cite{Cu2}. 
According to this, two valence bands near the Fermi level for
NaCo$_2$O$_4$ are well compared with those of the Ce-based compounds. 
One is the $a_{1g}+e_{g}$ band responsible for the electric conduction, 
corresponding to the $sp$ conduction band.
The other is the $a_{1g}$ band responsible for the large density of states, 
corresponding to the Ce $4f$ band.
Unlike the Ce-based compound, both bands cross the Fermi level 
to form two kinds of the Fermi surface.
Therefore, the rapid decrease in $m$ accompanied by the rapid 
decrease in $n$ can occur, when the Fermi surface of the $a_{1g}$ band 
disappears (or decouples with the $a_{1g}+e_{g}$ band)
upon the Pd substitution.
A possible candidate for the disappearance is a pseudogap opening.
The Cu substitution induces the spin-density-wave like transition 
at 22 K, below which the $a_{1g}$ Fermi surface seems to be gapped \cite{Cu2}.

Finally, we will briefly comment on some remaining issues. 
(i) Thermoelectric power factor $S^2/\rho$, 
a measure of thermoelectric performance, is proportional to $m/n$,
according to  equations (\ref{r}) and (\ref{s}).
In this case, good conduction adversely affects 
the thermoelectric properties.
In fact, the Pd substitution decreases 
the thermoelectric performance of NaCo$_2$O$_4$.
(ii) The magnetoresistance of the Pd-substituted samples is positive.
We previously reported that NaCo$_2$O$_4$ shows negative
magnetoresistance, which is attributed to 
the pseudogap in the $a_{1g}$ band \cite{terasaki2}.
In this sense the positive magnetoresistance 
is consistent with the disappearance of the $a_{1g}$ band,
but the field dependence is too complicated to analyze.
The angular dependence/anisotropy should be measured by using 
single crystals. Unfortunately
we have not yet succeeded in preparing Pd-substituted single crystals.
(iii) We failed to address how Pd modifies 
the electronic states  at a microscopic level. 
Our measurement is limited to the transport properties, 
and a microscopic probe such as photoemission is necessary.
Nevertheless we expect the small scattering cross section for Pd.
Probably Pd exists as Pd$^{2+}$ with the highest occupied orbital 
of $d_{z^2}$,
which is orthogonal to the valence bands of NaCo$_2$O$_4$ of $t_{2g}$.
(iv) We do not understand why the sign of $S$ and $R_H$ differs.
A number of materials show different signs of $R_H$ and $S$,
most of which are due to the multi-bands.
In the present case, however, we assumed that the single band gives 
the positive $S$ and the negative $R_H$.
A similar sign difference is seen in the Kondo semiconductor CeNiSn 
below the pseudogap temperature  \cite{takaba_rh,takaba_s},
and high-temperature superconductors \cite{terasaki3}.
In general, the signs of $S$ and $R_H$ are determined through
different averaging of the energy bands, 
which can be different even in a single band \cite{allen}.

\section{Summary}
In summary, we prepared a set of polycrystalline samples of 
Na$_{1.2}$Co$_{2-x}$Pd$_x$O$_4$ ($x$~=~0,~0.05,~0.1 and 0.2). The substituted Pd
decreases both the resistivity and the Seebeck coefficient, and increases 
the absolute value of the Hall coefficient. 
These results indicate that Pd decreases the carrier concentration 
and the effective mass significantly, 
while it does not influence the scattering time. 
Though microscopic details are still unknown, 
the present study reinforces our scenario that 
interplay between the two bands is a key to the high thermoelectricity
in NaCo$_2$O$_4$, just as in the case of heavy-fermion systems.

\section*{Acknowledgements}
The authors would like to appreciate K. Takahata for fruitful discussions 
and valuable comments. We would also like to thank W. Kobayashi and A. Satake 
for collaboration.


\end{document}